%% file: q_met.tex
\documentclass[twocolumn,secnumarabic,amssymb, nobibnotes, aps, pra]{revtex4-2}

\usepackage{graphicx}
\usepackage{bm}
\usepackage[T1]{fontenc}
\usepackage{color}
\usepackage{comment}
\usepackage{braket} 
\usepackage{amsmath}
\usepackage{amssymb}
\usepackage{mathtools} 
\usepackage{eqnarray,amsmath}
\usepackage{hyperref}

\usepackage{graphicx} 
\usepackage{xcolor}
\usepackage{epstopdf}
\usepackage[normalem]{ulem}

\begin{document}

\title{Distributed quantum sensing with optical lattices}

\begin{abstract}
In distributed quantum sensing the correlations between multiple modes, typically of a photonic system, are utilized to enhance the measurement precision of an unknown parameter. In this work we investigate the metrological potential of a multi-mode, tilted Bose-Hubbard system and show that it can allow for parameter estimation at the Heisenberg limit of $(N(M-1)T)^{2}$, where $N$ is the number of particles, $M$ is the number of modes, and $T$ is the measurement time. The quadratic dependence on the number of modes can be used to increase the precision compared to typical metrological systems with two atomic modes only, and does not require correlations between different modes. We show that the limit can be reached by using an optimized initial state given as the superposition of all the atoms occupying the first and the last site. Subsequently, we present strategies that would allow to obtain quadratic dependence on $M$ of the Fisher information in a more realistic experimental setup. 
\end{abstract}

\author{Jose~Carlos~Pelayo}
\email{jose-pelayo@oist.jp}
\author{Karol~Gietka}
\author{Thomas~Busch}
\affiliation{Quantum Systems Unit, Okinawa Institute of Science and Technology Graduate University, Okinawa, Japan 904-0495}

\date{\today}
\maketitle


\section{Introduction}
The main aim of quantum metrology is to understand how utilizing quantum effects can enhance the estimation precision of a parameter beyond the classical limits \cite{intro_met1,intro_met2,intro_met3,intro_met4}.  The latter is given by the shot-noise limit (SNL), which restricts the estimation precision to scale as $\Delta \theta \sim  (N\nu)^{-\frac{1}{2}}$, where $\theta$ is a parameter one wishes to measure precisely, $N$ is the number of particles being measured, and $\nu$ is the number of measurement repetitions. Making use of entangled states, one can enhance the precision by a factor of $k^{-\frac{1}{2}}$, where $k$ is the number of entangled particles \cite{useful_entanglement} and in the limit of a maximally entangled many-body state, i.e.~when $k=N$, one can achieve an estimation precision which scales as $N^{-1} \nu^{-\frac{1}{2}}$ for a decoherence-free system. This is the so-called \textit{Heisenberg scaling}, which provides an improvement of $\sim N^{-\frac{1}{2}}$ over the SNL \cite{H_limit}. Metrological enhancement over the SNL has been demonstrated by utilizing entangled states such as squeezed states \cite{squeezing1,squeezing2}, \textit{NOON} states \cite{NOON_1}, and others \cite{H_limit,GHZ,GHZ_met}, and has been implemented using a variety of setups such as squeezed states interferometers \cite{squeezed_interfero, squeezed_interfero2}, cavity QED~\cite{gietka2019,coherenceqedmet_2019,atomscavitymet_2020}, ion traps~\cite{Dalvitions_2006,metrologyiontrap_2018,amrscienceiontrap_2021}, and distributed quantum networks~ \cite{distributed_qmet1,distributed_qmet2,distributed_qmet3,zhou2022}. In distributed quantum networks, a single-mode photonic input state is fed to a global beam splitter network, which consists of  $M$  modes. The output state is then correlated between all of the modes before the parameter is imprinted and then measured at each mode. In this protocol, the parameter estimation has been shown to scale as $M^{-1}$, which corresponds to Heisenberg-like scaling for the system. This demonstrates that aside from the scaling with the number of particles, one can also utilize scaling with the number of modes to further enhance the parameter estimation. 

In standard setups for distributed quantum networks for photons, multi-mode correlations can be readily created through the use of beam splitters. However, in recent years precise control of atomic systems has become possible as well through the advent of atom cooling, trapping, and engineering techniques \cite{laser_cool_trap,evap_cool,laser_cool_trap_2}. Despite this, atomic systems have received relatively little attention in the context of distributed quantum networks~\cite{kasevich_2022} and in this work we want to bridge that gap and explore the metrological applications of multi-mode cold atomic systems by investigating how one can enhance the estimation precision by manipulating the controllable parameters of atoms trapped in an optical lattice.  


\section{Heisenberg limit in multi-mode, multi-particle atomic systems}

The ultimate bound to the estimation precision for a parameter $\theta$ over all possible measurements is given by the quantum Cram\'er-Rao bound, $\Delta\theta \ge \Delta\theta_{\text{QCR}} = 1/\sqrt{\nu F(\theta)}$, where $F(\theta)$ is the quantum Fisher information \cite{FI,QCRB1}. Given a Hamiltonian $H_\theta$ that depends on the parameter $\theta$ and evolves in time according to $U_\theta = \exp{(-iH_\theta t)}$, one can determine the quantum Fisher information by introducing a local generator $\hat{h}_\theta = i(\partial_\theta U_\theta)U_\theta^\dagger$, which characterizes the sensitivity of a state $\rho_\theta$ to an infinitesimal change in $\theta$, from $\rho_\theta \rightarrow \rho_{\theta + d\theta}$. For a system with an initial state that is a pure state $|\psi\rangle$, the quantum Fisher information can then be written as $F(\theta) = 4\langle\psi|\Delta^2 \hat{h}_\theta |\psi\rangle$ \cite{FI, FI2} and it is maximized when the initial state is given by an optimal state of the form $|\psi_{\text{opt}}\rangle = \frac{1}{\sqrt{2}}(|h_{\text{max}}\rangle + |h_{\text{min}}\rangle)$. In this case one gets
\begin{equation}
    F_{\text{max}}(\theta) = (h_{\text{max}} - h_{\text{min}})^2,
    \label{eq:FI_max}
\end{equation}
\noindent
where $h_{\text{max}}$ and $h_{\text{min}}$ are the maximal and minimal eigenvalues of $\hat{h}_\theta$  associated with states $|h_{\mathrm{max}}\rangle$ and $|h_\mathrm{min}\rangle$, respectively~\cite{intro_met4}. In the case where the Hamiltonian is composed of non-commuting terms, a compact expression for the local generator may not be available. An alternative expression to the local generator can be written in terms of the eigenvalues $E_k$ and eigenvectors $|\phi_k \rangle$ of the Hamiltonian \cite{Q_met_general_param} given as
\begin{align}
        \hat{h}_\theta &= \hat{h}_\theta^{(L)} + \hat{h}_\theta^{(O)}, \label{eq:generator}\\ 
        \hat{h}_\theta^{(L)} &= t\sum_{k=1}^{n_s}\frac{\partial E_k}{\partial \theta}|\phi_k \rangle \langle \phi_k| \label{eq:generator_L},  \\
        \hat{h}_\theta^{(O)} &=  2\sum_{l \neq k}e^{-\frac{itE_{kl}}{2}}\sin\left(\frac{tE_{kl}}{2}\right)\langle \phi_l|\partial_\theta\phi_k \rangle|\phi_k \rangle \langle \phi_l|, \label{eq:generator_O}
\end{align}
where $E_{kl} = E_{k} - E_l$ and $n_s$ is the total number of states, while $\hat{h}_\theta^{(L)}$ and $\hat{h}_\theta^{(O)}$ are the linear and oscillating parts of the local generator, respectively. Note that we use the convention $\hbar = 1$ throughout the manuscript.

To describe a multi-particle, multi-mode atomic system let us first consider a general Hamiltonian of the form
\begin{equation}
    H = \gamma \sum_m m \hat{a}^\dagger_m \hat{a}_m \label{eq:gen_Ham},
\end{equation}
where $\gamma$ is the parameter we wish to measure precisely, $\hat{a}^\dagger_m$ and $\hat{a}_m$ are creation and annihilation operators, and the mode-label $m$ runs from $1$ to $M$. The quantum Fisher information for a system described by Hamiltonian~\eqref{eq:gen_Ham} is maximized by a state 
\begin{equation}
    |\psi_{\text{opt}}\rangle = \frac{1}{\sqrt{2}}(|N0...0\rangle + |0...0N\rangle),
    \label{eq:optimum_state}
\end{equation}
which is a superposition of all the atoms occupying the first and the last site (for the sake of brevity we will call this a generalized $NOON$ state), and can be calculated to be
\begin{equation}
    F_{\text{max}}(\gamma) = T^2(N(M-1))^2 = F_{\text{HL}} \label{eq:HS}, 
\end{equation}
where $T$ is the length of the time interval during which the information about the unknown parameter was being imprinted. This will be our definition of the Heisenberg limit throughout this work. In particular, we will consider a 1D lattice system in a uniform linear potential, where $M$ is the total number of lattice sites, such that the enhancement proportional to $(M-1)$ is reminiscent to the scaling obtained from distributed photonic networks where $M$ refers to the number of modes. However, in these photonic systems, the maximum quantum Fisher information only scales with $\sim \bar{n}M^2$, where $\bar{n}\equiv N/M$ is the average photon number per mode \cite{distributed_qmet1}. Thus, once the Fisher information is expressed in terms of total amount of photons, the quadratic dependence on the number of modes disappears.

In the following we propose a possible realization of such a system using cold atoms and explore different strategies that can take advantage of the presence of multiple modes. 

\begin{figure}[tb]
    \centering  
    \includegraphics[width=\linewidth]{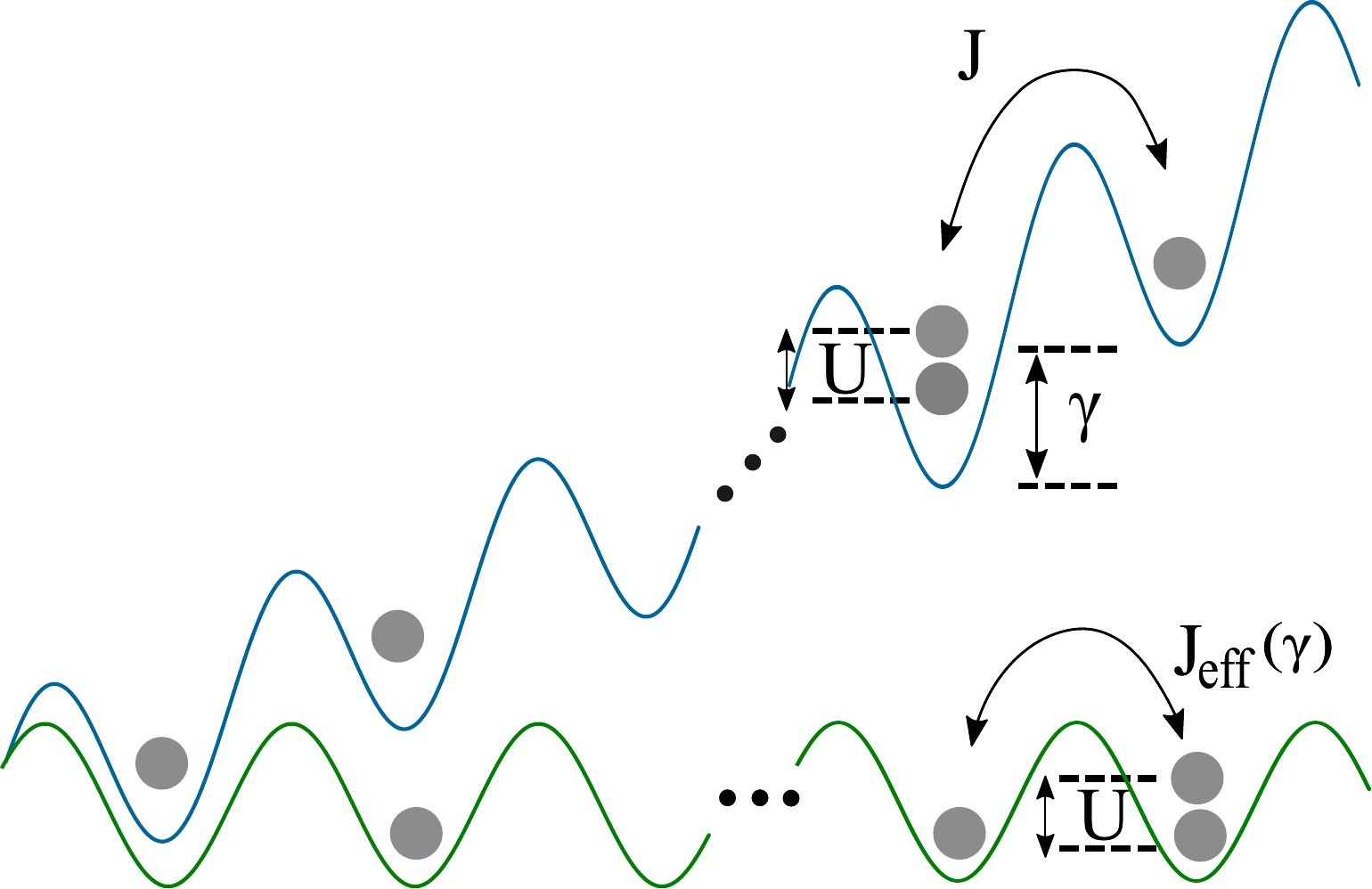}
    \caption{(top) Schematic of the TBH model. (bottom) Schematic of the driven TBH model in the effective time-independent picture as given in Eq.~\eqref{eq:DBH}. All parameters in Eq.~\eqref{eq:DBH}
    are scaled with respect to the tunneling coefficient, $J$, and are fixed for all simulations (except for $U$) with values: $J = 1$, $\gamma=33J$, $V_0 = 30.4J$, $\theta = \pi$, $\phi_{m+1} = \phi_m-\pi$, $\phi_0 = -\pi/2$. 
    }
    \label{fig:FINOON}

\end{figure}


\section{Tilted Bose-Hubbard Model}

A Hamiltonian with a term given in Eq.~\eqref{eq:gen_Ham} can be realised in a one-dimensional lattice system that is exposed to a uniform linear potential. Such a system is also known as a tilted Bose-Hubbard (TBH) model and can be described by

\begin{equation}
    H_{\text{TBH}} = -J\sum_{\langle i,j \rangle}  \hat{a}^\dagger_{i} \hat{a}_j + \gamma \sum_j j \hat{n}_j + \frac{U}{2}\sum_j \hat{n}_j(\hat{n}_j - 1) 
    \label{eq:TBH},
\end{equation}
where $J$ is the tunneling coefficient, $\langle\bullet\rangle$ denotes nearest neighbor sites, $U$ quantifies the on-site interaction strength between the particles, $\gamma$ is the strength of the tilt, and $\hat{n}_m = \hat{a}^\dagger_m \hat{a}_m$ is the number operator. A schematic of the system is shown in Fig.~\ref{fig:FINOON}(top) and we assume that we only have a finite number of sites, $j=1\dots M$. Although the TBH Hamiltonian contains additional terms when compared to Eq.~\eqref{eq:gen_Ham}, it is known that terms which are not dependent on the unknown parameter cannot alter the maximum attainable precision \cite{disturbance}. Thus, the Bose-Hubbard system in Eq.~\eqref{eq:TBH} will have a maximum quantum Fisher information that is still bounded by Eq.~\eqref{eq:HS}. 

The Hamiltonian in Eq.~\eqref{eq:TBH} can be realized using ultracold atoms trapped in a 1D optical lattice under the influence of a linear potential of strength $\gamma$, which, for example, can be gravitational or magnetic in nature \cite{H_H,harper_hofstadter2,gravimeter}. Therefore, a precise measurement of the parameter $\gamma$ would correspond to making a precise measurement of the field. The interaction strength $U$ can be tuned by employing Feshbach resonances \cite{Feshbach_M,Feshbach_O}, and is designed to be small relative to $\gamma$ such that excitations to higher bands are suppressed.

Assuming one can prepare the generalized $NOON$ state in Eq.~\eqref{eq:optimum_state} at $t=0$, the interaction can then be set to $U=0$, and the lattice depth to a value such that $J \ll \gamma $, which freezes the spatial dynamics. The remaining  dynamics is therefore purely in the phase difference between the two states in the superposition and is of the form $|\psi_{\text{opt}}(t)\rangle \sim \frac{1}{\sqrt{2}}(|N0...0\rangle + e^{-it\gamma N(M-1)}|0...0N\rangle)$ which is a state that yields $F_{\text{HL}}$ at all times. In this case all the information about the unknown parameter is stored in the relative phase between the two components of the wavefunction. It could be retrieved, for example, by performing an interference measurement, i.e., transferring the information stored in the relative phase to the occupation of each sites, which might be experimentally challenging. Furthermore, \textit{NOON} states are known to be very fragile against losses and thus difficult to prepare. Although several approaches have been suggested in the literature, especially for two-site systems, they usually suffer from having low fidelity as the number of particles is increased \cite{Weiss:09,Fogarty:13,NOON_state_repulse,Naldesi:22}.
We therefore explore in the following the prospects for metrology with a TBH model using a more realistic initial state. 

\begin{figure}[tb]
    \centering  
    \includegraphics[width=\linewidth]{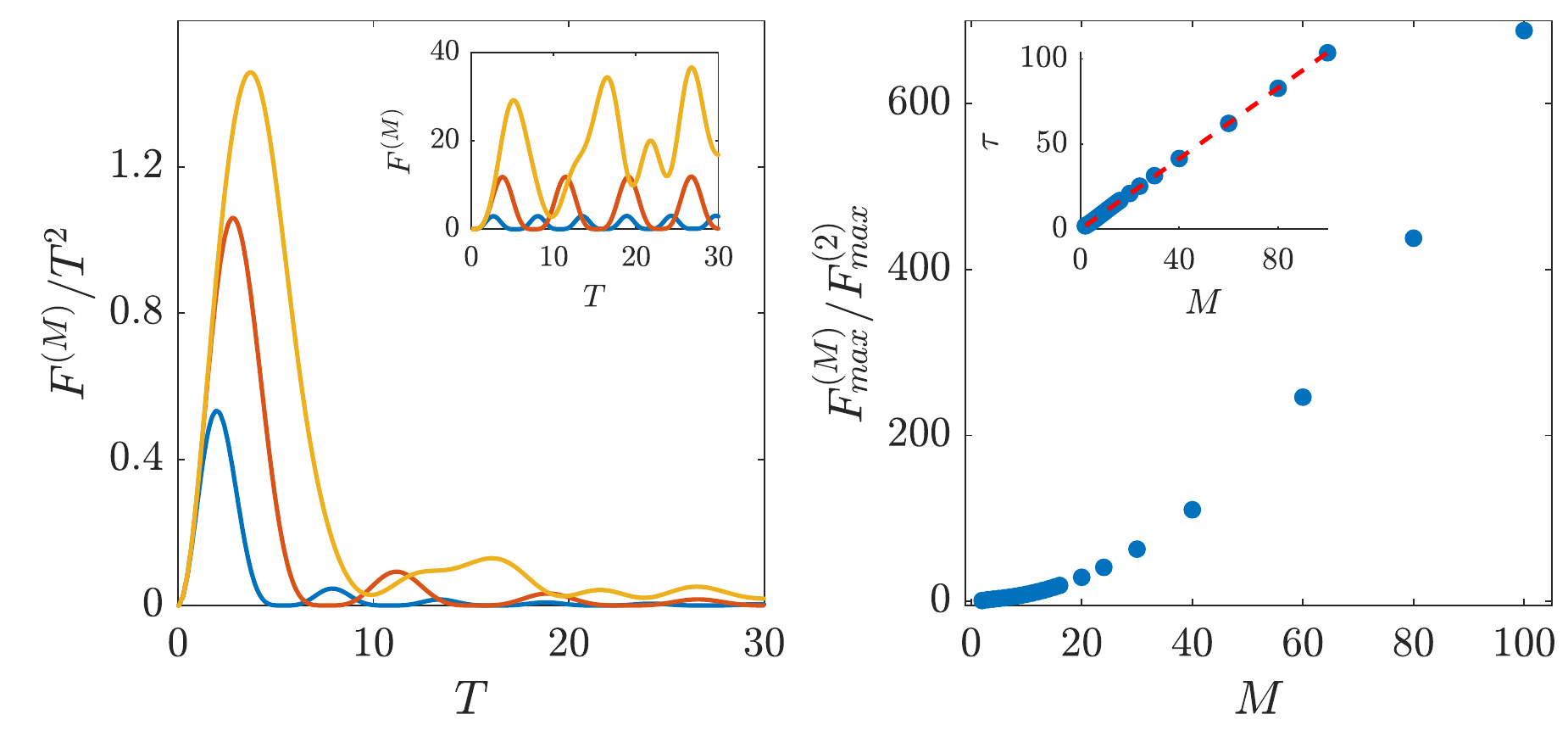}
    \put(-230,125){\textcolor{black}{(a)}}
    \put(-100,125){\textcolor{black}{(b)}}
    \caption{(a) Time dependence of the scaled $F^{(M)}$ for $M=\{2,3,4\}$ corresponding to solid lines $\{$blue, red, orange$\}$, respectively (the unscaled  $F^{(M)}$ is shown in the inset). (b) Growth of $F^{(M)}_{\text{max}}$ scaled by $F^{(M=2)}_{\text{max}}$ with $M$. The inset shows the linear scaling of $\tau$ with respect to $M$ (see text for details).}
    \label{fig:FIFock}
\end{figure}


\section{Fisher information with an initial Fock state} 
Let us start by considering an initial state where the particles are all placed in the  lowest energy site of the lattice 
\begin{equation}
    |\psi_{\text{Fock}}\rangle = |N0...0\rangle.
\end{equation}
\noindent

We first investigate the case where interactions are switched off and show later that when $U>0$, additional improvement to the quantum Fisher information can be observed due to the correlations introduced by the interactions. 

Given that $\gamma \gg J$, the spatial dynamics is frozen as before and since $4\langle\psi_{\text{Fock}}|\Delta^2 \hat{h}_\theta |\psi_{\text{Fock}}\rangle=0$, the Fisher information will be fixed to this value unless one restores the spatial dynamics. To introduce atomic dynamics to the system that depends on $\gamma$, we therefore consider a periodic drive with frequency $\omega = \gamma$, which involves the knowledge about the unknown parameter.
Such an approach would require, for example, an adaptive protocol where the knowledge about $\omega$ would be updated with every round of the protocol~\cite{wiseman_milburn_2009,2011Adaptivemetrology,2017Adaptivemetrologymarkovian,PRXQuantum.2.040301}. The driven Hamiltonian can then be written as
\begin{equation}
    H_{\text{DBH}} = H_{\text{TBH}} + V_0\sum_m \hat{n}_m \sin{\left(\omega t + \phi_m + \frac{\theta}{2}\right) \label{eq:DBH}},
\end{equation}
where $V_0$ is the driving amplitude, $\phi$ is a site-dependent phase, and $\theta$ is a constant phase. A schematic of the driven system in the effective time-independent picture is shown in Fig.~\ref{fig:FINOON}(bottom). These kinds of driving terms can be experimentally realized by an off-resonant laser-assisted tunneling scheme~\cite{harper_hofstadter1, harper_hofstadter2}. In order to suppress decoherence due to particle loss into the higher bands, we focus on the limit $\gamma \gg U$~\cite{Floquet_general,FT_bukov}  
and compute the resulting quantum Fisher information \cite{FI_eq2} of the system in Eq.~\eqref{eq:DBH} using 
\begin{equation}
     F(\gamma) = 4\left(\langle\partial_\gamma\psi_{\gamma}|\partial_\gamma\psi_{\gamma}\rangle - \left| \langle\psi_{\gamma}|\partial_\gamma\psi_{\gamma}\rangle \right|^2 \right).
     \label{eq:FI_time}
\end{equation}
\noindent
Fig.~\ref{fig:FIFock}(a) shows the evolution of the quantum Fisher information $F^{(M)}$ using the initial state $|\psi_{\text{Fock}}\rangle$ in the non-interacting driven  TBH model, where $M$ denotes the total number of lattice sites. Here, $F^{(M)}$ is scaled by $T^2$ and the number of particles is set to $N=1$ as it only has a linear contribution to the quantum Fisher information in the case of $U=0$, that is, $F^{(M,N)} = NF^{(M)}$. The inset of Fig.~\ref{fig:FIFock}(a) shows the oscillating behavior of the unscaled $F^{(M)}$ with $T$, and thus in the vicinity of  the first peak, the $F^{(M)}$ starts to scale poorly compared to $T^2$. 
This means that we only need to consider the dynamics up until the first peak of $F^{(M)}/T^2$ and we denote this peak as $F^{(M)}_{\text{max}} = \text{max}(F^{(M)}/T^2)$. Defining the time at which $F^{(M)}_{\text{max}} $ is attained as $\tau$, one can see in the inset of Fig.~\ref{fig:FIFock}(b) that it has a linear dependence on $M$, since more modes increases the time after which the state is transferred to the other end of the lattice. The quantum Fisher information enhancement relative to a two-level system, $F^{(M)}_{\text{max}}/F^{(M=2)}_{\text{max}}$, is shown in Fig.~\ref{fig:FIFock}(b) and clearly shows a quadratic dependence on $M$ for larger $M$. This illustrates that one can indeed make use of a larger system size in order to increase the quantum Fisher information even without introducing non-classical correlations. 

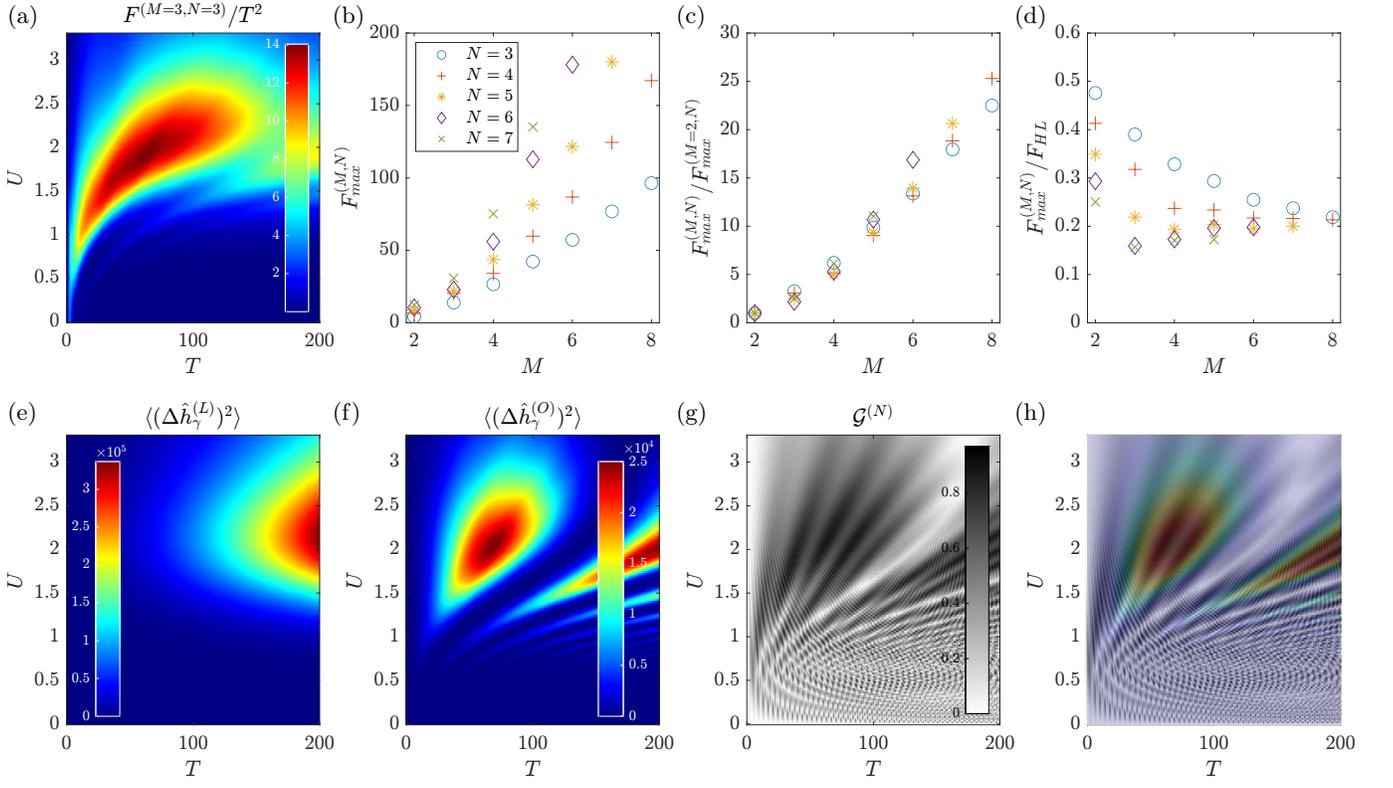
\begin{figure*}[tb]
    \centering  
    \includegraphics[width=\textwidth]{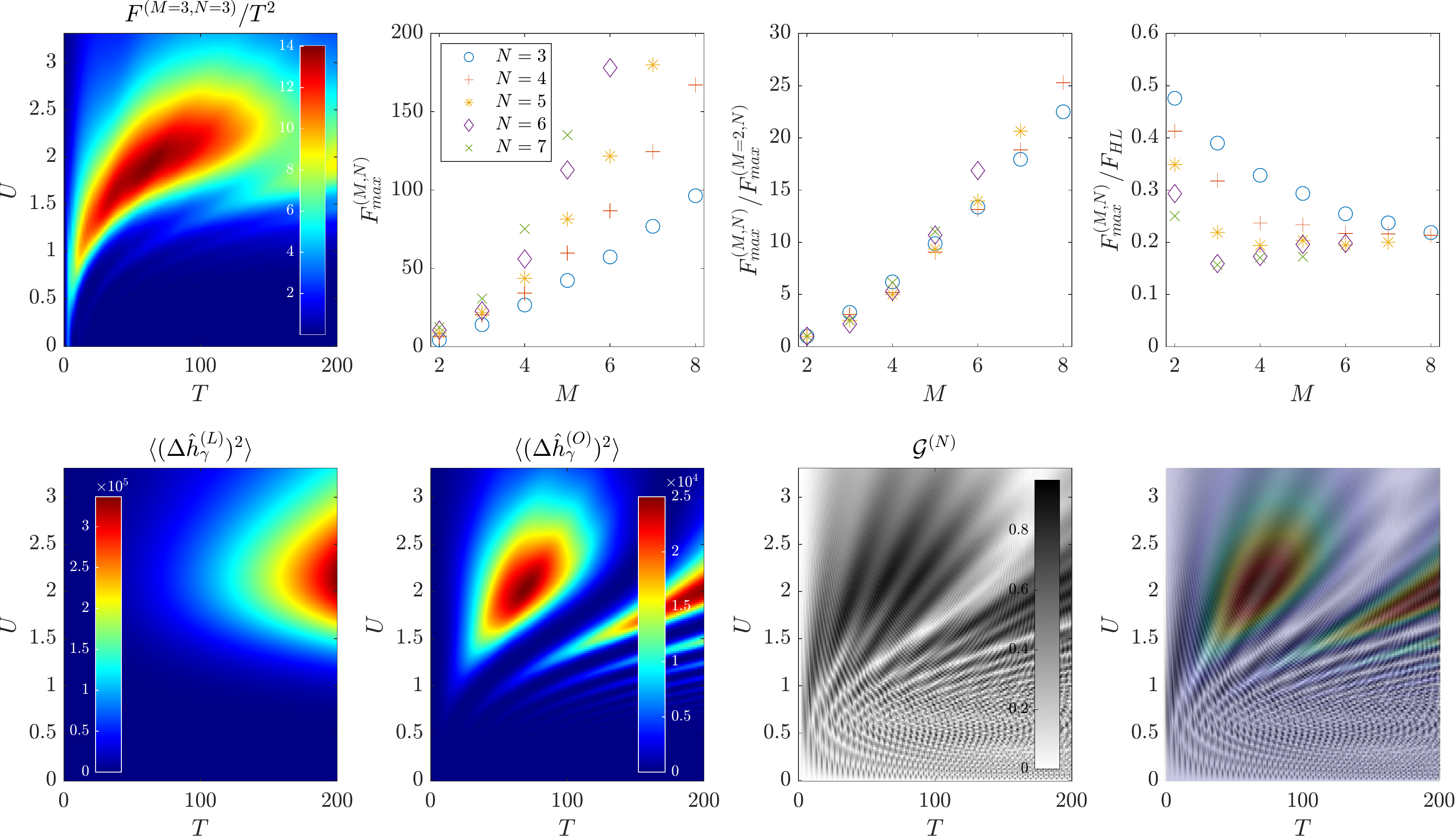}
    \put(-510,285){\textcolor{black}{(a)}}
    \put(-387,285){\textcolor{black}{(b)}}
    \put(-257,285){\textcolor{black}{(c)}}
    \put(-129,285){\textcolor{black}{(d)}}
    \put(-510,135){\textcolor{black}{(e)}}
    \put(-387,135){\textcolor{black}{(f)}}
    \put(-257,135){\textcolor{black}{(g)}}
    \put(-129,135){\textcolor{black}{(h)}}
    \caption{(a) Density plot of the scaled $F^{(M,N)}$ as a function of $T$ and $U$ for $N=M=3$. (b) Growth of $F^{(M,N)}_{\text{max}}$ with $M$ for $N=\{3,4,5,6,7\}$. The legend applies for plots (b)-(d). (c) $F^{(M,N)}_{\text{max}}$ scaled by $F^{(M=2,N)}_{\text{max}}$ and (d) scaled by the Heisenberg limit. Shown in (c) and (d) are all the data points that we were able to numerically obtain. (e) and (f) show the variance of $\hat{h}_\gamma^{(L)}$ and $\hat{h}_\gamma^{(O)}$, respectively. In (g) the correlation function, $\mathcal{G}^{(N)}$, is calculated for varying $U$ and $T$ and (h) shows the variance of $\hat{h}_\gamma^{(O)}$ overlaid by the correlation function. (See text for details).}
    \label{fig:FIFock2}
\end{figure*}


\section{Effect of interactions}
Next, we will consider an interacting system in which the additional non-classical correlations can be created that should improve the quantum Fisher information. The initial state is the same as in the previous section and we imprint the information about the unknown parameter and create the correlations at the same time~\cite{Hayes_2018,2021understandingCQM}. A representative surface plot of the evolution of $F^{(M,N)}/T^2$ with varying interaction strength $U$ for $N=M=3$ is shown in Fig.~\ref{fig:FIFock2}(a). Similar to the non-interacting case, we also observe a sinusoidal-like evolution of $F^{(M,N)}$ therefore we limit our interest up to the time $\tau$ when the first peak of $F^{(M,N)}/T^2$ appears. What is notable, however, is that a certain value of the interaction strength exists, which we will call $\bar{U}$, where the increase in $F^{(M,N)}$ over the non-interacting case is maximal. For the parameters used in Fig.~\ref{fig:FIFock2}(a)  this corresponds to $\bar{U}\approx1.92J$ and in general we observe an increase of $F^{(M,N)}_{\text{max}}$ as $N$ and $M$ increase (see Fig.~\ref{fig:FIFock2}(b)). When compared to a two-level system $F^{(M=2,N)}_{\text{max}}$, we still observe an enhancement as $M$ increases,  however, increasing the number of particles may not always yield a larger enhancement as shown in Fig.~\ref{fig:FIFock2}(c). Finally, when compared to the Heisenberg limit, the two-level system yields a stronger enhancement over any $M>2$ (for small $M$) and we again observe that the increase in $N$ does not always provide a larger enhancement as can be seen in Fig.~\ref{fig:FIFock2}(d). While it might look like that for increasing number of modes $F^{(M,N)}_{\text{max}}/F_{HL}\rightarrow \sim 0.2$, our limited computational resources currently do not allow us to explore this.%

To better understand the behavior of the quantum Fisher information $F^{(M,N)}$, we take the high-frequency approximation \cite{FT_dalibard_goldman,FT_dalibard_goldman_reso,Floquet_general,FT_bukov} of the driven, tilted Bose-Hubbard model up to first-order in $1/\omega$, which leads to an effective, time-independent description of the system given as,

\begin{align}
        H_{\text{eff}} =& -J_{F}\sum_{j}  (\hat{a}^\dagger_{j+1} \hat{a}_j e^{-i\phi_j} + h.c.) + \frac{U}{2}\sum_j \hat{n}_j(\hat{n}_j - 1)\nonumber \\ 
        & + K\left(\frac{1}{\omega}\right)\sum_j \left(\hat{n}_{j+1} - \hat{n}_j \right) + \mathcal{O}\left( \frac{1}{\omega^2}  \right)
    \label{eq:H_eff}
\end{align}
\noindent
where $J_F=J\mathcal{J}_1(2V/\omega)$  is the renormalized tunneling coefficient and $\mathcal{J}_1(x)$ is a Bessel function of the first kind. Since the effective Hamiltonian in Eq.~\eqref{eq:H_eff} is time-independent, the Fisher information can now be calculated from the local generator $\hat{h}_\gamma$ using Eq.~\eqref{eq:generator} and we look at the contributions from the linear part $\hat{h}_\gamma^{(L)}$ and the oscillating part $\hat{h}_\gamma^{(O)}$ separately. The respective variances are plotted in Figs.~\ref{fig:FIFock2}(e) and (f) and one can see  that $\langle(\Delta\hat{h}_\gamma^{(L)})^2\rangle$ dominates over $\langle(\Delta\hat{h}_\gamma^{(O)})^2\rangle$ at long times due to its quadratic dependence on time. The sinusoidal-like behavior that is seen in the inset of Fig.~\ref{fig:FIFock}(a) originates from $\langle(\Delta\hat{h}_\gamma^{(O)})^2\rangle$ and $\tau$ sets the timescale for the appearance of the first maximum of this term. The dependence of these generators to the eigenstates of $H_{\text{eff}}$ is discussed in the Appendix.
    
Finally, to illustrate that this enhancement of the quantum Fisher information is related to an increase in non-classical correlations, we make use of the $N$-th order correlation function given as \cite{NOON_state_correlator}
\begin{equation}
     \mathcal{G}^{(N)} = \left| \frac{1}{C}\langle \psi(t)| \left(\hat{a}^\dagger_M \hat{a}_1 \right)^N | \psi(t) \rangle \right|,
     \label{eq:correltion_function}
\end{equation}
\noindent
where $C = N!/2$ is a normalization constant. $\mathcal{G}^{(N)}$ quantifies the $N$-particle correlation between the two outermost sites, $j=1,M$. One can show that for a generalized $NOON$ state, $|\psi_{\text{opt}}(t)\rangle$, the correlation function is $\mathcal{G}^{(N)} = 1$ at all times. On the other hand, for an initial Fock state, $|\psi_{\text{Fock}}\rangle$, this is not the case. A surface plot of $\mathcal{G}^{(N)}$ as a function of $(T,U)$ is shown in Fig~\ref{fig:FIFock2}(g) for $|\psi_{\text{Fock}}\rangle$ and $M=N=3$. One can see that $\mathcal{G}^{(N)}$ has the same qualitative features as $\langle(\Delta\hat{h}_\gamma^{(O)})^2\rangle$ and by superimposing $\langle(\Delta\hat{h}_\gamma^{(O)})^2\rangle$ on $\mathcal{G}^{(N)}$ (see Fig.~\ref{fig:FIFock2}(h)) one can see that the maxima of $\langle(\Delta\hat{h}_\gamma^{(O)})^2\rangle$ lies in the region of large $\mathcal{G}^{(N)}$. This comparison however does not always hold true as can be seen for small $U$, when the correlator $\mathcal{G}^{(N)}$ suggests that there can be large correlation, but the quantum Fisher information is relatively small. However, we have confirmed for all combinations of $N=\{3,4,5\}$ and $M=\{3,4\}$ the regions of large $\langle(\Delta\hat{h}_\gamma^{(O)})^2\rangle$ also correspond to large correlations (not shown). As one goes to larger $N$ for a fixed $M$, the maximum correlation no longer approaches unity which could explain why the scaled $F^{(M,N)}_{\text{max}}$ in Fig~\ref{fig:FIFock2}(c)-(d) does not always increase even if $N$ is increased for a fixed $M$.


\section{Conclusions}
In this work, we have investigated the use of a  multi-mode atomic system in the context of distributed quantum sensing. We have shown that the driven, tilted Bose-Hubbard model can make use of the additional degree of freedom of the number of lattice sites or number of modes, $M$, in order to increase the quantum Fisher information of the system with respect to some unknown parameter. A generalized $NOON$ state maximizes the quantum Fisher information at all times when the spatial dynamics is frozen, however, extracting the information from such a state might require a complicated measurement procedure. An initial Fock state cannot saturate the Heisenberg limit but can still benefit from the quadratic scaling in $M$. In this case, the occupation of each site can be used as an optimal estimator. This quadratic scaling was made possible through the introduction of the periodic drive which translates the information about the parameter into the tunneling dynamics of the particles. On the other hand, for a distributed photonic network its quadratic scaling in $M$ becomes linear once the quantum Fisher information is expressed in terms of the total number of particles. Additionally, by introducing interactions to the system we have shown that parameter imprinting and creation of correlations can be achieved simultaneously in contrast to the distributed photonic networks where the creation of the correlated state is performed before the parameter imprinting.

We emphasize that the enhancement with respect to the number of modes is not an unexpected result. Suppose we have a harmonic oscillator where the energy spacing is $\omega_0$, and suppose further that there is a non-linear process that can drive the system from the ground state to the $M$-th level with energy $\omega_M = M\omega_0$, then using the error propagation formula we get that the uncertainty in $\omega_0$ is given by $\Delta\omega_0 = \Delta\omega_M / M$. This is the same enhancement we observe for the tilted, Bose-Hubbard model using the generalized $NOON$ state. The only difference here is that no non-linear process is used to couple neighboring modes and thus demonstrates the advantage of utilizing a ladder-like system with multiple modes in metrology. It is clear that the existence of the tilt (or any dispersion which is of the form $~m^\alpha$, where $\alpha \ge 1$ and $m$ is the mode index) is what allows the enhancement with the number of modes, therefore, other ladder-like systems can also exploit this enhancement. An example of such a system is the periodically forced Bose-Hubbard model \cite{periodically_forced} given as, $H_{\text{PF}} = H_0 + \Gamma\cos{(\omega t) \sum_j j\hat{n}_j} $ where the difference between this and the tilted Bose-Hubbard model is that the tilt is no longer linear but instead it is periodic in time. What is interesting here is that if we precisely measure $\omega$ instead of $\Gamma$, we find that the Heisenberg limit is $(\Gamma NM)^2T^4$, where we still have the enhancement in $N$ and $M$ but now the scaling with time is super-quadratic and can be achieved only if one incorporates an additional optimal control Hamiltonian \cite{FI_eq2}. These ladder-like systems provide an avenue to study quantum metrological setups which have a potential to further increase precision measurements.


\section{Acknowledgements}
The authors would like to acknowledge Lewis Ruks for fruitful discussions. 
This work was supported by the Okinawa Institute of Science and Technology Graduate University (OIST). The authors are also grateful for the the Scientific Computing and Data Analysis (SCDA) section of the Research Support Division at OIST.
\noindent


\appendix*
\section{Eigendecomposition of $H_{\text{eff}}$ \label{Appendix_more_discussions}}
\begin{figure}[tb]
    \centering  
    \includegraphics[width=\linewidth]{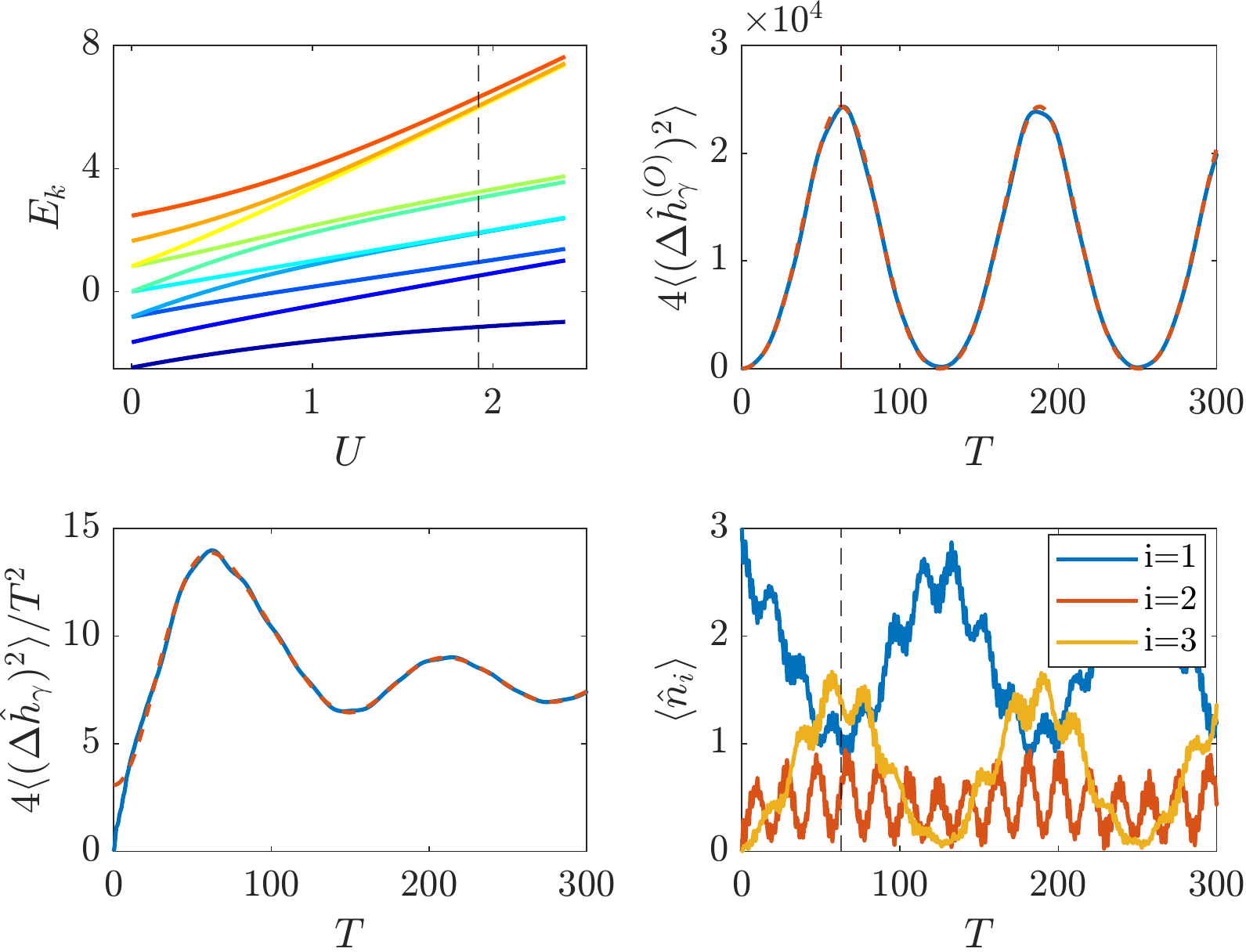}
    \put(-250,185){\textcolor{black}{(a)}}
    \put(-125,185){\textcolor{black}{(b)}}
    \put(-250,88){\textcolor{black}{(c)}}
    \put(-125,88){\textcolor{black}{(d)}}
    \caption{The plots are determined for the case $M=N=3$. (a) The eigenvalues $E_k$ are shown as a function of increasing interaction strength $U$. In (a) The vertical dashed line corresponds to $U = \bar{U}$ while in (b) and (d) the vertical dashed line corresponds to $T = \tau$. In (b) and (c) the variances $\langle(\Delta\hat{h}_\gamma^{(O)})^2\rangle$ and $\langle(\Delta\hat{h}_\gamma)^2\rangle/T^2$ as a function of $T$ are calculated using the full oscillating generator $\hat{h}_\gamma^{(O)}$ (blue solid line) and the approximate generator $\tilde{\hat{h}}_\gamma^{(O)}$ (orange dashed line). (d) The site occupation number $\langle\hat{n}_j\rangle$ is plotted as a function of $T$. Plots (b)-(d) are calculated at fixed $U = \bar{U}$. (See text for details)}
    \label{fig:eigensystem}
\end{figure}
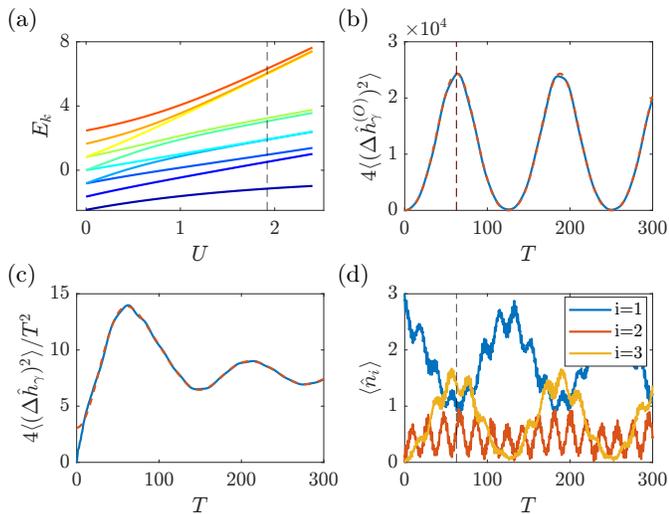

 Another perspective on the local generator and quantum Fisher information is to look at the eigenvalues $E_k$ and eigenfunctions  $|\phi_k \rangle$ of the effective Hamiltonian in Eq.~\eqref{eq:H_eff}. In Fig.~\ref{fig:eigensystem}(a) we plot the energy of each of the eigenfunctions as a function of increasing interactions $U$, where the vertical dashed-line corresponds to $U = \bar{U}$ at which we observe $F^{(M,N)}_{\text{max}}$. One can see immediately that at this critical value some of the eigenfunctions are close to degeneracy. This determines which terms in the sum in Eq.~\eqref{eq:generator_O} are dominant when calculating $\langle(\Delta\hat{h}_\gamma^{(O)})^2\rangle$ since it contains $\langle \phi_l|\partial_\gamma\phi_k \rangle = \frac{\langle \phi_l|\frac{\partial H_{\text{eff}}}{\partial \gamma}|\phi_k \rangle}{E_k - E_l}$. While in Fig.~\ref{fig:eigensystem}(a) we see that there are two pairs of eigenfunctions that are close to being degenerate, only one of these pairs actually gives a large contribution and corresponds to the 2nd and 3rd largest eigenstates. The reason for this is that these two states, $|\phi_{n_s - 1} \rangle$  and $|\phi_{n_s - 2} \rangle$, correspond to the ones with the largest overlap with the initial Fock state, $|\langle \phi_{n_s - 1}|\psi_{\text{Fock}}\rangle|^2 \approx 0.17$ and $|\langle \phi_{n_s - 2}|\psi_{\text{Fock}}\rangle|^2 \approx 0.72$, respectively. 
 The energy difference between this pair, $\Omega = E_{n_s - 1}-E_{n_s - 2}$, is then related to the oscillation frequency of $\langle(\Delta\hat{h}_\gamma^{(O)})^2\rangle$, and thus one can approximate $\tau$ as $\tau \sim \frac{\pi}{\Omega}$ (see Fig.~\ref{fig:eigensystem}(b)). The oscillating part of the local generator $\hat{h}_\gamma^{(O)}$ can then be approximated by just considering the states $|\phi_{n_s - 1} \rangle$  and $|\phi_{n_s - 2} \rangle$. We define this approximate generator as $\tilde{\hat{h}}_\gamma^{(O)}$ and its variance is also shown in Fig.~\ref{fig:eigensystem}(b). At long times this becomes a good approximation even when calculating the quantum Fisher information as shown in Fig.~\ref{fig:eigensystem}(c).  Finally, we determine the average occupation at each site $\langle\hat{n}_j\rangle$ as a function of $T$ for $U = \bar{U}$ (see Fig.~\ref{fig:eigensystem}(d)). Here we observe that at $T \sim \tau$ the average occupation between the first and last sites gets close which suggests that an observable such as $\sim\langle\hat{n}_1\rangle\langle\hat{n}_M\rangle$ can be a good estimator.
 
\input{q_met.bbl}

\end{document}

%% file: q_met.bbl
%